\documentclass[aps,prl,twocolumn,showcaps,superscriptaddress,showpacs,amsmath,amssymb]{revtex4}

\usepackage{graphicx}
\usepackage{dcolumn}
\usepackage{bm}
\usepackage[dvips]{epsfig}
\usepackage{color}
\usepackage{upgreek}
\usepackage{color}

\newcommand{\unit}[1]{\;\mathrm{#1}}

\newcommand{\kB}{\ensuremath{k_{\rm{B}}}}

\newcommand{\qdot}{\ensuremath{\dot{q}}}
\newcommand{\x}{\ensuremath{x}}
\newcommand{\xvec}{\ensuremath{\vec{\x}}}
\renewcommand{\v}{\ensuremath{\dot{x}}}
\newcommand{\vvec}{\ensuremath{\dot{\vec{x}}}}
\newcommand{\nois}{\ensuremath{\zeta}}
\newcommand{\noisvec}{\ensuremath{\boldsymbol \nois}}
\newcommand{\vs}{\ensuremath{v^{\rm s}}}

\newcommand{\ps}{\ensuremath{p^{\rm s}}}

\newcommand{\jsvec}{\ensuremath{\vec{j}^{\rm s}}}

\newcommand{\nusvec}{\ensuremath{{\boldsymbol\nu}^{\rm s}}}

\newcommand{\nablavec}{\ensuremath{\boldsymbol\nabla}}
\newcommand{\mob}{\ensuremath{\mu}}

\newcommand{\Fvec}{\ensuremath{\vec{F}}}

\renewcommand{\vec}[1]{\mathbf #1}

\newcommand{\mean}[1]{\langle #1 \rangle}

\begin{document}

% \title{Efficient method to measure dissipation in colloidal systems}
% \title{Measuring heat production efficiently on the level of trajectories}
\title{Noninvasive measurement of dissipation in colloidal systems}

\author{B.~Lander}
\affiliation{II. Institut f\"ur Theoretische Physik, Universit\"at Stuttgart, Pfaffenwaldring 57, 
70550 Stuttgart, Germany}
\author{J.~Mehl}
\affiliation{2. Physikalisches Institut, Universit\"at Stuttgart, Pfaffenwaldring 57, 70569 
Stuttgart, Germany}
\author{V.~Blickle}
\affiliation{2. Physikalisches Institut, Universit\"at Stuttgart, Pfaffenwaldring 57, 70569 
Stuttgart, Germany}
\affiliation{Max-Planck-Institute for Intelligent Systems, Heisenbergstrasse 3, 70569 Stuttgart, 
Germany}
\author{C.~Bechinger}
\affiliation{2. Physikalisches Institut, Universit\"at Stuttgart, Pfaffenwaldring 57, 70569 
Stuttgart, Germany}
\affiliation{Max-Planck-Institute for Intelligent Systems, Heisenbergstrasse 3, 70569 Stuttgart, 
Germany}
\author{U.~Seifert}
\affiliation{II. Institut f\"ur Theoretische Physik, Universit\"at Stuttgart, Pfaffenwaldring 57, 
70550 Stuttgart, Germany}


% ---------- Abstract ----------

\begin{abstract}
According to Harada and Sasa [Phys. Rev. Lett. \textbf{95}, 130602 (2005)], heat production
generated in a non-equilibrium steady state can be inferred from measuring response and correlation
functions. In many colloidal systems, however, it is a nontrivial task to determine response
functions, whereas details about spatial steady state trajectories are easily accessible. Using
a simple conditional averaging procedure, we show how this fact can be exploited to reliably
evaluate average heat production. We test this method using Brownian dynamics simulations, and
apply it to experimental data of an interacting driven colloidal system.
\end{abstract}

\maketitle

{\it Introduction.---}The phenomenon of dissipation distinguishes equilibrium from non-equilibrium
systems. For macroscopic systems, dissipation can be inferred either directly through measuring
temperature changes or from the known external work applied to a system. At least for
non-equilibrium steady states (NESSs) the latter is equal to the dissipation. For small systems such
as colloidal particles or molecular motors, measuring dissipation is highly nontrivial. Calorimetric
methods on the single particle or molecule level fail due to the tiny values of the heat generated
by single degrees of freedom. In principle, the framework of stochastic 
thermodynamics~\cite{seif07,seki10, jarz11} allows to apply the first law to phenomena on this
scale from which the exchanged heat could be extracted if both, the externally applied work to drive
the system and the internal energy change could be measured, where the latter vanishes in a NESS. In
practice, however, knowing the external force applied to a colloidal particle by a laser-field
beyond the paradigmatic harmonic trap~\cite{wang02} is quite a challenge~\cite{blic06, jop08}.
Likewise, measuring directly the amount of adenosine triphosphate (ATP) molecules hydrolyzed by a
single molecular motor is impossible. Thus, the ingredients of using the first law to infer
dissipation, in general, are not directly accessible.

Harada and Sasa suggested an exact relation quantifying heat production in terms of the violation
of the fluctuation-dissipation relation (FDR) for systems with a single degree of freedom obeying
overdamped Langevin dynamics~\cite{hara05}. Extensions to many-body systems both
for over- and underdamped Langevin dynamics~\cite{hara06}, as well as Hamiltonian
systems~\cite{tera05} followed shortly after. Their results have been experimentally tested for
an optically driven colloidal system~\cite{toya07} and applied to gain information about
the non-equilibrium energetics of F$_1$-ATPase~\cite{toya10}. Further generalizations and
experiments applying this technique comprise Langevin systems including memory~\cite{deut06,
toya08}, quantum Langevin dynamics~\cite{sait08a}, and FDR violations involving field
variables~\cite{hara09}. For molecular motors connected to a colloidal probe, another approach has
been put forward, from which information about dissipation can be deduced~\cite{toya12, zimm12}.

Specifically, the Harada-Sasa relation expresses the average heat production rate as
\begin{equation}
 \label{eq:harada-sasa}
  \mean{\qdot} = \sum\limits_{i=1}^n \mob^{-1}_i \left\{ (v^s_i)^2 + \int_{-\infty}^{\infty} \left[
\tilde{C}_{ii}(\omega) -
2T\tilde{R}_{ii}'(\omega) \right] \frac{d\omega}{2\pi} \right\},
\end{equation}
where Boltzmann's constant is set to unity, $T$ is the temperature of the surrounding heat bath and
$\mob_i$ is the mobility of the $i$th of the $n$ degrees of freedom. The response function
\begin{equation}
  R_{ij}(t) \equiv \frac{\delta\mean{\v_i(t)}}{\delta f_j(0)}
\end{equation}
quantifies the change of the velocity $\v_i$ to a force perturbation applied to the $j$th degree of
freedom. Furthermore,  
\begin{equation}
  C_{ij}(t) \equiv \mean{ (\v_i(t) - \vs_i) (\v_j(0) - \vs_j)}
\end{equation}
is the velocity autocorrelation function, and $\tilde{g}(\omega) \equiv
\int_{-\infty}^{\infty} g(t)\exp{(i\omega t)} dt /(2\pi)$ is the Fourier transform of an
arbitrary function $g(t)$. The prime in Eq.~\eqref{eq:harada-sasa} denotes the real part of a
complex-valued function.

This exact relation yields insight into the origins of FDR violations and can be used readily
if response functions are accessible. Determining response functions for driven colloidal systems,
however, often poses a nontrivial and time-consuming task due to the need to perturb the system
from its steady state. The perturbation must be sufficiently small to stay within the linear
response regime and has to be applied to each degree of freedom separately.
Determining response functions in frequency space requires separate measurements for each frequency,
which must be done up to sufficiently high frequencies to make sure that the integral in
Eq.~\eqref{eq:harada-sasa} converges. If determined in temporal space, the system must be observed
during its relaxation, which excludes the possibility to average over the time coordinate, thus
substantially increasing the statistical effort.

In experiments and simulations, good statistics for trajectories in a NESS is readily
available. In the following, we will employ a conditional averaging procedure on these trajectories
from which the mean local velocity field can be obtained. Combined with the measured stationary
distribution, this field then yields the average heat production rate. This procedure avoids
difficulties arising from the application of external perturbations and reduces the effort to create
the statistics needed. It thus complements the approach given by Harada and Sasa by making the
average heat production rate easily accessible from steady state trajectories. Likewise, it does not
require any knowledge about the applied forces or interactions within the system.

{\it Theory.---}We introduce this method for systems whose dynamics is governed by a set of coupled 
overdamped Langevin equations
\begin{equation}
  \label{eq:Langevin}
  \vvec =  \mob \Fvec(\xvec) + \noisvec,
\end{equation}
with the coordinate $\xvec = (x_1, \dots, x_n)$, and the force $\Fvec$ composed of conservative and
nonconservative contributions. Interactions with the surrounding solvent are modeled by Gaussian
white noise $\noisvec$ with zero mean $\mean{\noisvec(t)} = 0$ and correlations
\begin{equation}
 \label{eq:white_noise}
  \mean{\noisvec(t) \noisvec^T(t')} = 2 \mob T \delta(t-t').
\end{equation}
The stationary probability distribution function $\ps(\xvec)$ follows from the 
Smoluchowski equation according to
\begin{equation}
 \label{eq:smol}
  0 = - \nablavec \cdot \jsvec(\xvec)
\end{equation}
with the probability current
\begin{equation}
 \label{eq:jsvec}
  \jsvec(\xvec) \equiv \mob \Fvec(\xvec) \ps(\xvec) - \mob T \nablavec \ps(\xvec).
\end{equation}
The closely related mean local velocity is defined as the conditional average of the fluctuating
velocity $\vvec$ at position $\xvec$,
\begin{multline}
  \label{eq:def:nus}
  \nusvec(\xvec) \equiv
%  \mean{ \vvec | \xvec } =\\
  \lim\limits_{\Delta t \to 0}
  \mean{\xvec(t+\Delta t) - \xvec(t-\Delta t) | \xvec(t) = \xvec} / (2\Delta t).
\end{multline}
Here, the Stratonovich convention has to be employed, thus evaluating the spatial variable in
mid-step position. For the remainder of this Rapid Communication, we will always assume this convention.
This relation enables us to evaluate the mean local velocity from NESS trajectories. Carrying out the
average analytically, one obtains~\cite{seif07}
\begin{equation}
 \label{eq:nus}
 \nusvec(\xvec) = \jsvec(\xvec)/\ps(\xvec) = \mob \Fvec(\xvec) - \mob T \nablavec
\ln{\ps(\xvec)}.
\end{equation}

The average heat production rate along a stochastic trajectory $\xvec(t)$ is given by~\cite{seki10}
\begin{equation}
 \label{eq:def:avg_heat}
  \mean{\qdot} \equiv \mean{\Fvec(\xvec(t)) \cdot \vvec(t)}.
\end{equation}
Since $\xvec$ is a stochastically fluctuating quantity some care needs to be taken while
averaging. In the Stratonovich scheme, one has~\cite{seif07}
\begin{equation}
 \label{eq:qdot_fluct}
  \mean{\qdot} = \mean{ \vec{F}(\xvec) \cdot \nusvec(\xvec) }.
\end{equation}
Now, all fluctuating quantities have been replaced. Using Eq.~\eqref{eq:nus} we substitute
$\vec{F}$ and obtain
\begin{equation}
 \label{eq:qdot_new:nonNESS}
  \mean{\qdot} =  \mob^{-1} \int d\xvec\, \nusvec(\xvec)^2 \ps(\xvec) 
    + T \int d\xvec\, \nusvec(\xvec) \cdot \nablavec \ps(\xvec).
\end{equation}
The second integral on the right-hand side vanishes, since
\begin{multline}
  \int d\xvec\, \nusvec(\xvec) \cdot \nablavec \ps(\xvec) = \int d\xvec\, \jsvec(\xvec) \cdot
\nablavec\ln{\ps(\xvec)}   \\= -\int d\xvec\, [\nablavec\cdot\jsvec(\xvec)] \ln{\ps(\xvec)} = 0.
\end{multline}
Here, we have used Eq.~\eqref{eq:smol} after a partial integration. The boundary term vanishes
due to the periodicity of $\jsvec$ and $\ps$. Therefore, the average heat production rate reads
\begin{equation}
  \label{eq:qdot_new}
  \mean{\qdot} = \mob^{-1} \int d\xvec\, \nusvec(\xvec)^2 \ps(\xvec) =
\mob^{-1}\mean{\nusvec(\xvec)^2}.
\end{equation}
Given a long trajectory $\xvec(t)$, we can determine the stationary distribution $\ps$ and via
Eq.~\eqref{eq:def:nus} also the mean local velocity $\nusvec$. Therefore, we are able to
evaluate the average heat production rate in the system simply by recording particle trajectories
in the NESS.

In the following, we will first illustrate the validity and usefulness of the method by using
Brownian dynamics simulations to model a driven system consisting of two coupled
colloidal particles. Second, we apply the method to data for the experimental system
and determine the average heat production rate as a function of the coupling strength.

% ---------- two ring system in simulation ----------
\begin{figure}[h]
 \centering
 \includegraphics[width=\linewidth]{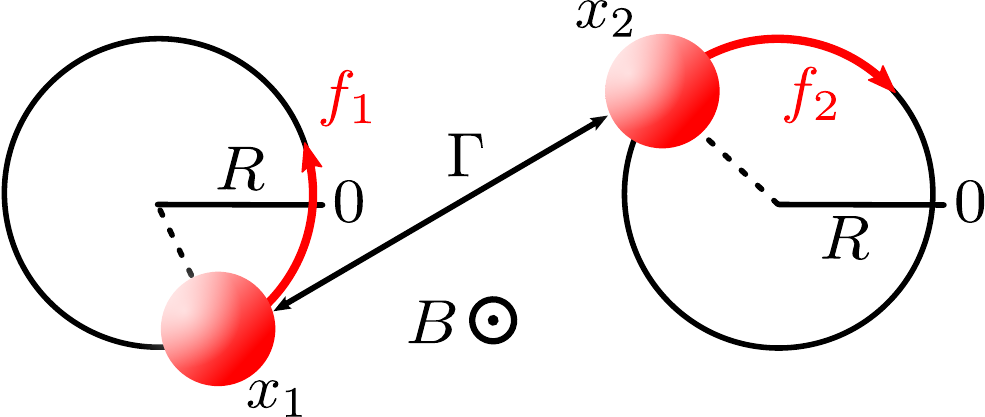}
 \caption{Schematic representation of the system: Two paramagnetic colloidal particles driven
along two rings of radius $R$ by constant forces $f_i$. The position $x_i$ of the $i$th particle is
the arc length measured in the counterclockwise direction. The magnetic field applied perpendicular to
the rings induces a magnetic moment in each particle. The strength of the resulting repulsive
interaction is quantified by the dimensionless plasma parameter $\Gamma$.
\label{fig:Fig1}}
\end{figure}

{\it Simulation.}---We use Brownian dynamics simulations to test this method
for the following, experimentally accessible~\cite{mehl12}, colloidal system. Two paramagnetic
colloidal particles are driven on two nonoverlapping rings of radius $R$ [see Fig.~\ref{fig:Fig1}].
By applying a small homogeneous magnetic field $B$ normal to the plane containing the rings, the
particles acquire parallel magnetic dipole moments $m \approx \alpha B$ with $\alpha \simeq 5.9
\times 10^{-12}\unit{Am^2/T}$. The resulting repulsive interaction is described by the potential
$W(\xvec) = \mu_0 m^2/ (4\pi r^3(\xvec))$, where $\mu_0$ is the magnetic constant and $r(\xvec)$
the distance between the particles. In order to quantify the interaction strength, we introduce the
dimensionless plasma parameter $\Gamma\equiv\Delta W / T$, where $\Delta W$ is the difference
between the maximum and minimum in the interaction energy. Apart from the interaction, the total
force acting on the particle on the $i$th ring,
\begin{equation}
  \label{eq:2dsys:force}
  F_i(\xvec) =  f_i -\partial_{x_i} \left(V_i(x_i) + W(\xvec) \right),
\end{equation}
contains the sinusoidal potential $V_i = A_i \sin(x_i /R - \phi_i)$ of amplitude $A_i$ and phase
shift $\phi_i$, and the constant driving force $f_i$.
Using these forces, the dynamics of the system is given by the Langevin
equation~\eqref{eq:Langevin}. The equations of motion are integrated via a stochastic
Runge-Kutta algorithm~\cite{kloeden} using a time step of $\Delta t = 0.0001\unit{(\upmu
m)^2/}(\mob {\rm T})$.

\begin{figure}
 \centering
 \includegraphics[width=\linewidth]{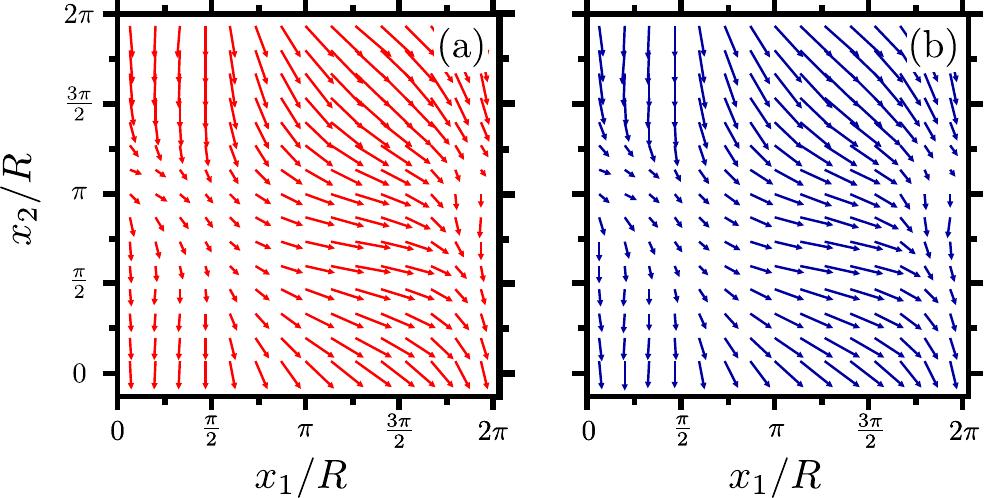}
 \caption{Comparison between the force fields $\Fvec(\xvec)$ for the two-ring system obtained
(a) analytically and (b) from NESS trajectories via Eq.~\eqref{eq:nus} for parameter set $I$ [see
Table~\ref{tab:parameters}].
\label{fig:Fig2}}
\end{figure}

The most crucial point of our method is the determination of the mean local velocity $\nusvec$ and the
probability distribution function $\ps$. In order to check this point, we proceed as follows.
First, we compute the force field $\Fvec$ analytically, which is possible since in this simulation
all potential parameters are known in detail. Second, we compute $\Fvec$ via Eq.~\eqref{eq:nus} by
evaluating both $\nusvec$ and the gradient field of $\ps$ from the simulated data. In
Fig.~\ref{fig:Fig2}, we compare the differently obtained force fields and find very good agreement.

We calculate the average heat production rate using two different methods: (A) We use the trajectory-based
method introduced in this Rapid Communication, i.~e.,we determine $\nusvec$ and $\ps$ from the simulated
trajectories and use Eq.~\eqref{eq:qdot_new} to obtain $\mean{\qdot}$. (B) As a test, we determine
$\mean{\Fvec \cdot\vvec}$ along the complete trajectory for each particle. The sum then yields the true
average heat production rate in this simulation run. Of course, this procedure is possible only if details
of the interaction are known, as is the case in these Brownian dynamics simulations.

\begin{figure}
  \includegraphics[width=\linewidth]{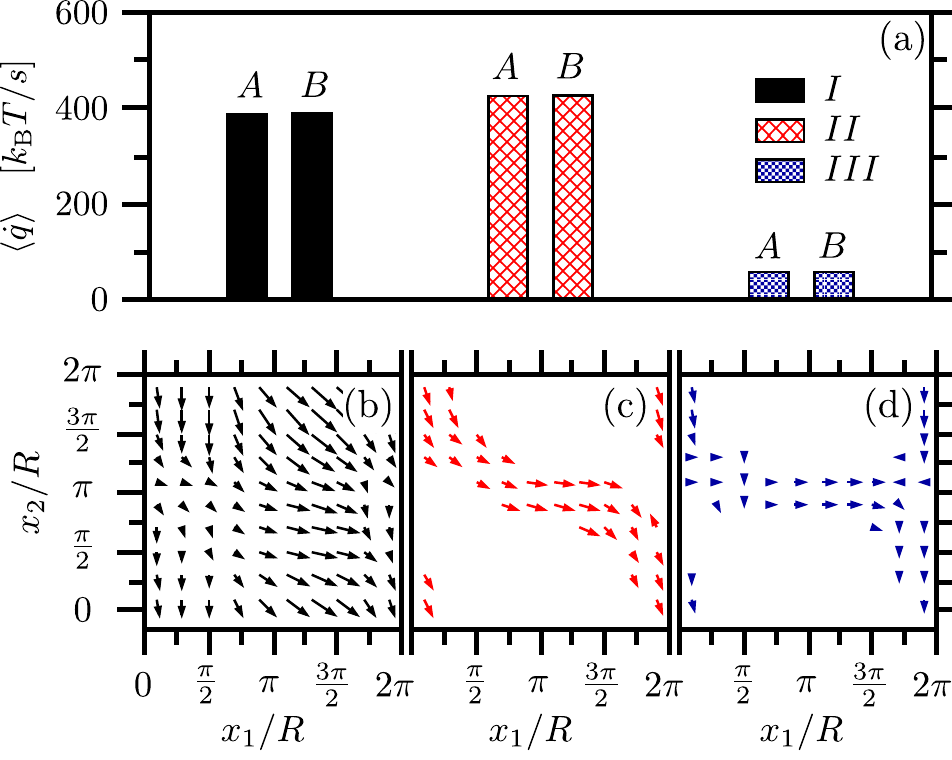}
  \caption{(a): Average heat production rates $\mean{\qdot}$ determined from methods (A) and (B) as
described in the text for three different parameter sets $I$, $II$, and $III$ [
see Table~\ref{tab:parameters}]. To relate $\mean{\qdot}$ to an experimental scale we choose $a =
5.2\unit{\upmu m}$. Error bars are smaller than 1\%. (b)-(d): Corresponding evaluated mean local
velocity fields $\nusvec(\xvec)$. Regions without arrows are rarely visited by the system.
\label{fig:Fig3}}
\end{figure}

In Fig.~\ref{fig:Fig3}, we present a comparison between the differently evaluated average heat
production rates determined for three different parameter sets [see Table~\ref{tab:parameters}].

\begin{table}
  \centering
  \begin{tabular}{lccc}
      \hline\hline
						&	$I$		&	$II$	&$III$\\
      \hline
      $\Gamma$			&		$310.0$	&	$670.0$	& $310.0$\\
      $f_1 \unit{[\kB T / \upmu m]}$	&	$50.0$	&	$56.0$	& $25.0$\\
      $f_2 \unit{[\kB T / \upmu m]}$	&	$-65.0$	&	$-54.0$	& $-35.0$\\
      $A_1 \unit{[\kB T]}$		&	$225.0$	&	$175.0$	&$175.0$\\
      $A_2 \unit{[\kB T]}$		&	$175.0$	&	$170.0$	&$175.0$\\
      $\phi_1$			&	$-\pi/5$	&	$-\pi/5$	&$-\pi/5$\\
      $\phi_2$			&	$-3\pi/5$	&	$-3\pi/5$	&$-3\pi/5$\\
      $R \unit{[\upmu m]}$	&	$3.5$	&	$3.5$	&$3.5$\\
      center-to-center distance $[\upmu \mathrm{m}]$  &	$17.0$	& $17.0$	& $17.0$\\
      \hline\hline
  \end{tabular}
  \caption{Parameter sets $I$, $II$, $III$. \label{tab:parameters}}
\end{table}

We find very good agreement, supporting the consistency between the present approach (A) and the
reference result (B). The most critical point in the application of method (A) is the determination
of $\nusvec$ in those phase-space regions where the system is found
with low probability. The average heat production rate $\mean{\qdot}$, however, is computed from an
integral [see Eq.~\eqref{eq:qdot_new}] in which $(\nusvec)^2$ enters weighted by $\ps$. Since in rarely visited regions this weight is small, their contributions to $\mean{\qdot}$
should play a minor role, provided that $\nusvec$ stays finite. This is the case in our system, and
should hold for many other well-behaved systems as well. Therefore, to obtain averaged quantities
such as $\mean{\qdot}$, only regions in configuration space of substantial weight must be sampled
accurately. Regions of minor probability will only add small-sized corrections to the result.

% ---------- experimental system ----------

{\it Experiment.}--- Having demonstrated its validity, we now apply the method to the 
experimental system~\cite{mehl12}. Two paramagnetic colloidal particles with diameters
$a=5.2\unit{\upmu m}$ are driven by a scanning laser beam~\cite{fauc95,mehl10} along
two rings with radii $R=3.5\unit{\upmu m}$ and a center-to-center distance of $17\unit{\upmu m}$.
We checked for hydrodynamic interactions by measuring the mean velocity of a particle driven along
a circle with constant potential in the presence of another particle at minimal distance $d$. For
$d\geqslant9\unit{\upmu m}$, we do not find any deviation from the $d\to\infty$ limit.
  We thus conclude that hydrodynamic interactions are negligible in this case~\footnote{
  Additionally, we have estimated the strength of the hydrodynamic coupling in relation to the
  dipole interaction up to the Rotne-Prager level. Comparing the contributions of the different
  types of interaction in the range of angles where the particles come closest, we find that 
  hydrodynamic contributions are $\lesssim20\%$ of the dipolar contribution for the smallest
  non-zero  $\Gamma$ used and smaller by at  least one order of magnitude  for all others.
}.
Using digital video microscopy, we track the NESS trajectories $x_1(t)$ and $x_2(t)$ with a
spatial and temporal resolution of $20\unit{nm}$ and $25\unit{ms}$, respectively. We determine the
average heat production rate for different magnetic fields, $B\leqslant40\unit{mT}$, thus changing
the plasma parameter in the range of $0\leqslant\Gamma\leqslant1100$. 

\begin{figure}
  \includegraphics[width=\linewidth]{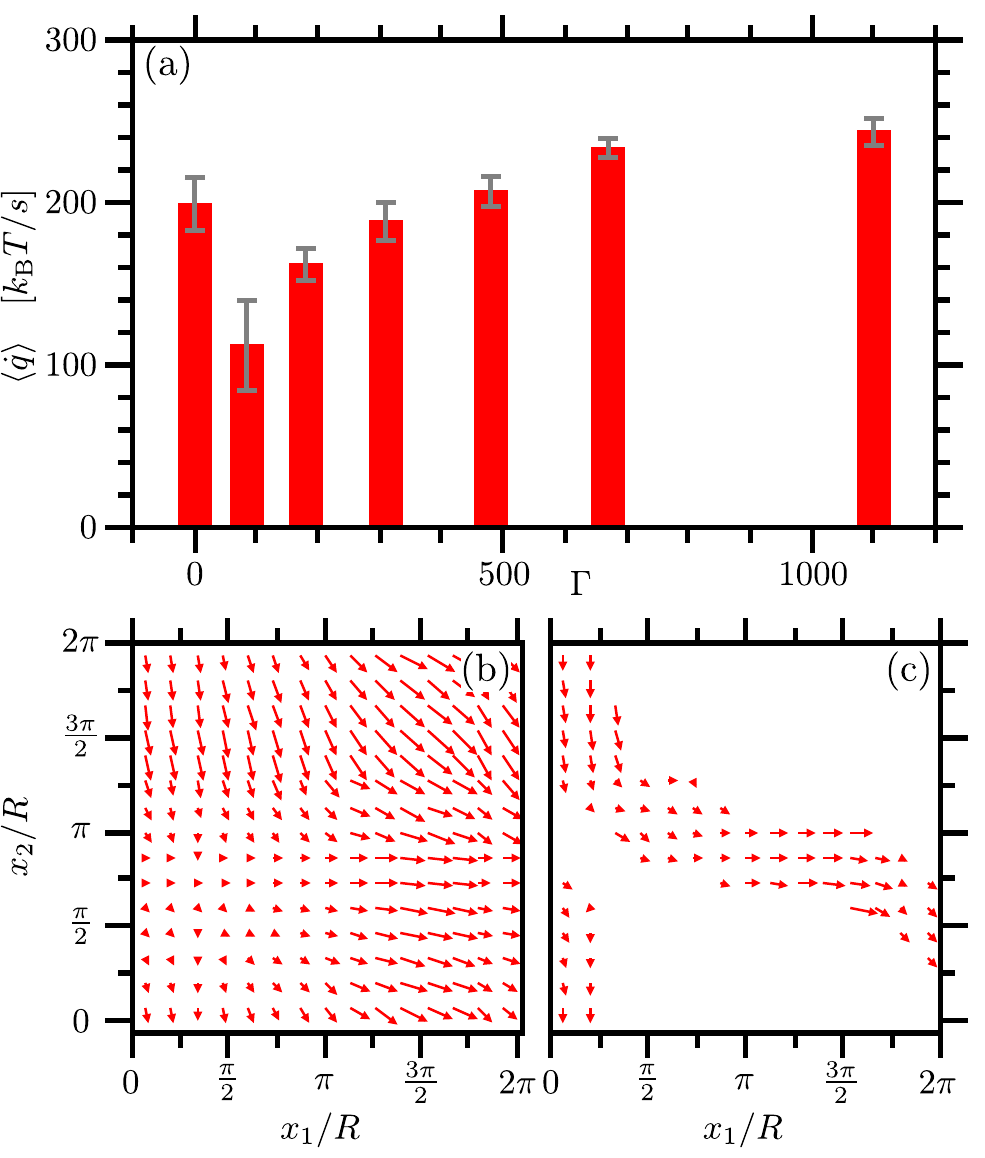}
  \caption{(a): Average heat production rates $\mean{\qdot}$ for different plasma parameters 
$\Gamma$. The parameters for the NESSs are $f_1 = 56\unit{\kB T / \upmu m}$, $A_1 = 175\unit{\kB
T}$, and $f_2 = -51\unit{\kB T / \upmu \rm m}$, $A_1 = 164\unit{\kB T}$ obtained by the method
presented in Ref.~\cite{blic07a}. Error bars are obtained by evaluating the data set for four parts
of equal size and computing the standard deviation. (b)-(c): Corresponding mean local velocity
fields $\nusvec(\xvec)$ determined from the NESS trajectories for plasma parameters (b)~$\Gamma = 0$,
and (c)~$\Gamma = 1150$, respectively. \label{fig:Fig4}}
\end{figure}

In the absence of coupling, the particles move independently along their tilted potentials with a 
mean circulation time of $12\unit{s}$. This motion leads on average to a heat production rate of 
$\mean{\qdot}=\mu^{-1}\mean{\nusvec(\xvec)^2}=199\unit{\kB T/ s}$ with $\mu^{-1}=3\pi\eta a$ the
bare mobility of one particle and $\eta$ the viscosity of the solvent. Under strong coupling 
conditions, $\mean{\qdot}$ reaches higher values [see
Fig.~\ref{fig:Fig4}~(a)]. This fact becomes obvious when focusing on the corresponding mean local
velocity fields $\nusvec$.
For $\Gamma=0$, $\nusvec$ shows a large region with very small
velocities [see Fig.~\ref{fig:Fig4}~(b)] corresponding to potential minima at $x_1/R=\pi/4$ and
$x_2/R=3\pi/4$. For $\Gamma\geqslant480$, the motion of the two particles synchronizes due to the
repulsive interaction as follows [see Fig.~\ref{fig:Fig4}~(c)]. While the first particle fluctuates
around its potential minimum, the second one completes one circulation. Arriving at
the point where the particles are closest, this particle pushes the first one over its barrier. Then, the roles switch and the first particle completes a circulation while the second
one fluctuates around its potential minimum. This effect shortens the time to overcome the potential
barrier considerably for both particles and results, on average, in a faster circulation, and thus a
larger $\mean{\qdot}$ than in the uncoupled case. In the intermediate regime, the particles may
still overtake one another. During the overtaking process, the faster particle hinders the other
one in overcoming the potential barrier, effectively slowing down its motion. This on average slower
motion produces less heat, and consequently $\mean{\qdot}$ is non-monotonic in $\Gamma$ [see
Fig.~\ref{fig:Fig4}~(a)]. We emphasize that we obtained robust results for $\mean{\qdot}$ for NESS
trajectories with a length of only $30\unit{min}$ corresponding to $150$ full revolutions of each
particle.

{\it Concluding perspectives.}---In this Rapid Communication, we have presented an approach to
determine the average heat production rate for colloidal systems in a NESS. While
the Harada-Sasa method quantifies dissipation via response and correlation functions, the present
method solely uses information obtained from steady state trajectories without the need to perturb
the system. Neither does the method require any information about the underlying potentials or
driving forces. Recording particle trajectories in the NESS is sufficient.
This approach thus constitutes a complementary way to measure dissipation, which is
easily implemented particularly in set-ups concerned with small colloidal systems.
By replacing the stationary distribution function and the stationary mean local velocity field by
their time-dependent counterparts~\cite{seif07}, a generalization of the present method should also
be applicable to relaxing and time-dependent systems.

% ---------- Bibliography ----------
% \bibliography{refs.bib}

\end{document}